\documentclass[11pt]{article}
\usepackage{natbib}
\usepackage{amssymb,amsmath,amsthm,latexsym,amsfonts,amscd,dsfont,enumerate}
\usepackage{a4wide,graphicx,tikz}

\usetikzlibrary{arrows,shapes,shadows,fit,backgrounds}
\tikzstyle{trader} = [circle, draw, top color=white, bottom color=blue!30, draw=blue!50!black!100, drop shadow, minimum height=4em]
\tikzstyle{bank} = [rectangle, draw, top color=white, bottom color=red!20, draw=red!50!black!100, drop shadow, rounded corners, minimum height=3em, text width=4em, text centered]
\tikzstyle{market} = [rectangle, draw, top color=white, bottom color=green!20, draw=green!50!black!100, drop shadow, rounded corners, minimum height=3em, text width=4em, text centered]
\tikzstyle{background} = [rectangle,fill=gray!10, inner sep=0.2cm, rounded corners=5mm]
\tikzstyle{line} = [draw, latex'-latex']
\tikzstyle{from} = [draw, latex'-]
\tikzstyle{to} = [draw, -latex']

\usepackage[]{hyperref}
 \hypersetup{
 colorlinks=true, 
 breaklinks=true, 
 urlcolor= blue, 
 linkcolor= blue, 
 bookmarksopen=true, 
 pdfauthor={Pallavicini, A., Brigo D.}
 }

\newtheorem{theorem}{Theorem}[section]
\newtheorem{remark}[theorem]{Remark}

\newcommand{\Ex}[2]{\mathbb{E}_{#1}\!\left[\,#2\,\right]}
\newcommand{\ExT}[3]{\mathbb{E}_{#1}^{#2}\!\left[\,#3\,\right]}

\newcommand{\Px}[1]{\mathbb{P}\left\{\,#1\,\right\}}

\newcommand{\ind}[1]{1_{\{#1\}}}

\newcommand{\cov}[3]{{\rm Cov}_{#1}^{#2}\!\left[\,#3\,\right]}

\newcommand{\lgd}{\mbox{L{\tiny GD}}}

\newcommand{\Eq}[1]{{\[{#1}\]}}
\newcommand{\beq}[0]{\begin{equation}}
\newcommand{\eeq}[0]{\end{equation}}
\newcommand{\beqn}[0]{\begin{equation*}}
\newcommand{\eeqn}[0]{\end{equation*}}
\newcommand{\balign}[0]{\begin{aligned}}
\newcommand{\ealign}[0]{\end{aligned}}

\title{Interest-Rate Modelling in Collateralized Markets:\\ multiple curves, credit-liquidity effects, CCPs}
\author{
Andrea Pallavicini\thanks{Imperial College London and Banca IMI Milan, {\tt a.pallavicini@imperial.ac.uk}}
\ \ \
Damiano Brigo\thanks{Imperial College London, {\tt damiano.brigo@imperial.ac.uk}}
}
\date{
\small First Version: October 1, 2012.  This version: \today
}

\begin{document}

\maketitle

\begin{abstract}
The market practice of extrapolating different term structures from different instruments lacks a rigorous justification in terms of cash flows structure and market observables. In this paper, we integrate our previous consistent theory for pricing under credit, collateral and funding risks into term structure modelling, integrating the origination of different term structures with such effects. Under a number of assumptions on collateralization, wrong-way risk, gap risk, credit valuation adjustments and funding effects, including the treasury operational model, and via an immersion hypothesis, we are able to derive a synthetic master equation for the multiple term structure dynamics that integrates multiple curves with credit/funding adjustments. 
\end{abstract}

{\bf JEL classification code: G13. \\ \indent AMS classification codes: 60J75, 91B70}

\medskip

{\bf Keywords:} Yield Curve Dynamics, Multiple Curve Framework, HJM Framework, Interest Rate Derivatives, Basis Swaps, Counterparty Credit Risk, Liquidity Risk, Funding Costs, Central Clearing Counterparties.

\newpage
{\small \tableofcontents}
\vfill
{\footnotesize \noindent The opinions here expressed  are solely those of the authors and do not represent in any way those of their employers.}
\newpage

\pagestyle{myheadings} \markboth{}{{\footnotesize  A. Pallavicini, D. Brigo, Multi-Curve models explained via Credit, Collateral and Funding Liquidity}}

\section{Introduction}
\label{sec:introduction}

Starting from summer 2007, with the spreading of the credit crunch, market quotes of forward rates and zero-coupon bonds began to violate standard no-arbitrage relationships. This was partly due to the liquidity crisis affecting credit lines, and to the possibility of a systemic break-down triggered by increased counterparty credit risk. Indeed, credit risk is only one facet of the problem, since the crisis started as a funding liquidity crisis, as shown for example by \cite{Tapking2009}, and it continued as a credit crisis following a typical spiral pattern as described in \cite{Brunnermeier2009}.

This has been the most dramatic signal for inadequacy of standard financial modelling based on idealized assumptions on risk-free rates and on unrestricted access to funding instruments. Removing or relaxing such assumptions opens the door to financial models able to analyze the inner mechanics of a deal: collateral rules, funding policies, close-out procedures, market fees, among others. Such features require to be included in our pricing framework.

A few papers in the financial literature have tried either to re-define the theory from scratch or to extend the standard framework to the new world. The works of \cite{Crepey2011,Crepey2012a,Crepey2012b} open the path towards a pricing theory based on different bank accounts, accruing at different rates, and representing the different cash sources one may have at her disposal to implement the hedging strategies, the funding policies and the margining rules. We cite also the works by \cite{Piterbarg2010,Piterbarg2012}, which focus on perfectly collateralized deals and try to reformulate the basic Black Scholes theory, and the works of \cite{BurgardKjaer2011a,BurgardKjaer2011b}. On the other hand, the works of \cite{Perini2011,Perini2012} stress the fact that all updated features can be included in terms of modified payoffs rather than through the need of a new and somehow ad-hoc pricing theory. In particular, \cite{Perini2011,Perini2012} recognize that a financial contract must include hedging, funding and 
margining fees as specific and precisely defined additional cash flows.  

In this paper we try to distill the above contributions to highlight the common ingredients that a new pricing framework should include to be effective in a world where credit and liquidity issues cannot be disregarded. We formulate a term structure theory that is based on the new pricing framework and that leads quite naturally to multiple curves. The cornerstone of our analysis will be market procedures and market observables. We avoid employing pricing rules that are not supported by market strategies. For example, funding requires a counterparty willing to lend cash to the investor, hedging requires trading only the instruments offered by the market, and so on. On the other hand, we wish to maintain the discussion as simple as possible and to use well developed pricing and hedging concepts as much as possible, so that we still use some theoretical quantities, such as a risk-neutral measure and a risk free rate. However, we use such unobservables only as instrumental variables and we check that our final 
results are independent of them. Our final multiple-curves term structure theory, rigorously integrated in terms of credit, collateral and funding effects, will be based only on market observables and market procedures.

In particular, we wish to analyze concrete pricing cases such as partial collateralized deals and contracts cleared by Central Clearing Counterparties (CCP), which is a very sensitive topic following pressure from Dodd Frank and EMIR/CRD4. See for instance \cite{Arnsdorf2011}, \cite{Pirrong2011}, \cite{Rama2011}, and \cite{Heller2012}.

\medskip

We now move to describing more in detail the content of the paper. 

In Section \ref{sec:pricing} we present our pricing framework along with realistic approximations.

We start by pointing out that the classical relationship between LIBOR forward rates and zero coupon bonds is not working anymore. In particular, interpreting the zero-coupon curves as risk-free rate curves is no longer tenable. Likewise, interpreting the LIBOR rates as the simply compounded rates that underlie risk-free one-period swaps does not work anymore. To include risks that are now heavily affecting interest rates, we design a theory that includes explicit cash flows accounting for default close-out and also for costs of funding for the hedging portfolio, and costs of funding for collateral (margining).

In particular, we assume that {\emph{LIBOR is now a rate assigned by the market and not derived by risk-free forward rate agreements or risk-free one-period swaps}}.

By including credit, collateral and funding effects in valuation we obtain the master equation seen earlier in \cite{Perini2011,Perini2012}. In the specific case where collateral is a fraction of current all-inclusive mark-to-market, we obtain a simpler pricing equation based on a hedge-funding-fees equivalent measure and on a generalized dividend process inclusive of default risk, treasury funding and collateral costs. We do not consider collateral gap risk for interest-rate products, as such risks are not essential for this asset class, as described in \cite{BrigoCapponiPallaviciniPapatheodorou} (contrary to credit derivatives, as seen in \cite{BrigoCapponiPallavicini}).

In Section \ref{sec:ir} we apply these approximations to the money market, in order to evaluate collateralized interest-rate derivatives.

We look at defining new building blocks that will replace the old unobservables-based ones (such as risk-free rate zero-coupon bonds, risk-free one-period swap rates, etc). Such new instruments are based on the collateral rate, which is an observable rate, since it is contractually defined by the CSA as the rate to be used in the margining procedure. The collateralized zero-coupon bond can be proxied by one-period Overnight Indexed Swaps (or by quantities bootstrapped from multi-period OIS) when we accept to approximate a daily compounded rate with continuously compounded rates. We then define fair OIS rates at inception in terms of collateralized zero coupon bonds.

Thus, {\emph{this setup allows us to define new LIBOR forward rates as equilibrium rates in collateralized one-period swaps.}

The resulting rates depend on the collateralized coupon-bearing bonds. We may also define a collateral based forward measure where  LIBOR forward rates are expected values of future realized LIBOR rates. We then hint at the development of a market model theory for the new collateral-inclusive LIBOR forward rates and to a forward rates theory for the OIS-based instantaneous forward rates.

This is where the multiple curve picture finally shows up: {\emph{we have a curve with LIBOR based forward rates, that are collateral adjusted expectation of LIBOR market rates we take as primitive rates from the market, and we have instantaneous forward rates that are OIS based rates.} OIS rates are driven by collateral fees, whereas LIBOR forward rates are driven both by collateral rates and by the primitive LIBOR market rates.

We conclude this section by introducing a dynamical multiple-curve model for OIS and LIBOR rates. We reformulate the parsimonious HJM model by \cite{Moreni2010,Moreni2012} under our new pricing framework.

In Section \ref{sec:funding} we focus on uncollateralized, partially-collateralized and over-collateralized contracts.

We remove perfect collateralization and adopt a partial one. This means that now we need to evaluate the pricing adjustment by including also the corrections coming from the treasury and hedging strategy funding costs.

In particular, we have that {\emph{LIBOR forward rates associated to partially collateralized one-period swap contracts acquire a covariance term that can be interpreted as a convexity adjustment.}

CCPs can be modelled in the above framework and, in particular, initial margins can be roughly modelled with a collateral fraction larger than unity. This leads to a generalization of the previous immersion-based general formula.

In Section~\ref{sec:modelling} we discuss how to model credit spreads, funding rates and specific assumptions on the fraction of mark-to-market covered by collateral.

Credit spreads should be calibrated via Credit Default Swaps or Defaultable Bonds. However, this should be a global calibration because CDS are collateralized and are in principle priced with the same general formula, inclusive of collateral and funding, that we use for all other deals. Bonds are not collateralized and are funded more heavily, so funding risk is also there. This global CDS or Bond calibration is not done usually even though interpreting CDS or Bonds as sources of pure credit risk calibration may lead to important errors, see for example \cite{Fujii2011CDS} or \cite{BrigoCapponiPallavicini}.

The term structure of funding rates depends on the funding policy and is model dependent. Stripping it directly from market liquid instruments is very difficult, especially because the interbank market is no longer (fully) representative of such costs. Collateral portfolios play now a key role too. Default intensities, collateral rates and liquidity bases will be key drivers of the funding spreads. Finally, we consider using Value-at-Risk measures to determine the effective fraction of mark-to-market we should hold as collateral, and introduce collateral haircuts.  

Our final interest-rate curves are consistently explained by such effects and based on market observables. 

In Section \ref{sec:conclusion} we conclude the discussion and list open points.

\section{Valuation under Credit, Collateral and Funding risk}
\label{sec:pricing}

After the onset of the crisis in 2007, all market instruments are quoted by taking into account, more or less implicitly, credit and funding liquidity adjustments. Hence we have to carefully check standard theoretical assumptions which often ignore credit and liquidity issues. We must go back to market data by limiting ourselves to price only derivative contracts we are able to replicate by means of market instruments. The comparison to market data and processes is the only means we have to validate our theoretical assumptions, so as to drop them if in contrast with observations.

Our general results are not linked to a particular asset class, but in this paper we focus as main example on interest-rate derivatives, as a paradigmatic case.

\subsection{The Credit Crunch}

Classical interest-rate models were formulated to satisfy no-arbitrage relationships by construction, which allow to hedge forward-rate agreements in terms of risk-free zero-coupon bonds.

Starting from summer 2007, with the spreading of the credit crunch, market quotes of forward rates and zero-coupon bonds began to violate usual no-arbitrage relationships. The main driver of such behavior was the liquidity crisis reducing the credit lines along with the fear of an imminent systemic break-down. As a result the impact of counterparty risk on market prices could not be considered negligible any more.

This is the first of many examples of relationships that broke down with the crisis. Assumptions and approximations stemming from pricing theory should be replaced by strategies implemented with market instruments. For instance, inclusion of credit valuation adjustments (CVA) for interest-rate instruments, such as those analyzed in \cite{BrigoPallavicini2007}, breaks the relationship between risk-free zero coupon bonds and LIBOR forward rates. Also, funding in domestic currency on different time horizons must include counterparty risk adjustments and liquidity issues, see \cite{Filipovic2012}, breaking again this relationship. We thus have
\Eq{
F_t(T_0,T_1) \neq \frac{1}{T_1-T_0} \left( \frac{P_t(T_0)}{P_t(T_1)} - 1 \right) \,,
}%
where $P_t(T)$ is a zero coupon bond price at time $t$ for maturity $T$, and $F$ is the related LIBOR forward rate. A direct consequence is the impossibility to describe all LIBOR rates in terms of a unique zero-coupon yield curve. Indeed, since 2009 and even earlier, we had  evidence that the money market for the Euro area was moving to a multi-curve setting. See \cite{Bianchetti2009}, \cite{Henrard2009}, \cite{Kijima2009}, \cite{Mercurio2009,Mercurio2010}, \cite{PallaviciniTarenghi}, and \cite{Fujii2011}.

A further example of evolving assumptions is given by the use of collateralized contracts. The growing attention on counterparty credit risk is transforming OTC derivatives money markets. An increasing number of derivative contracts is cleared by CCPs, while most of the remaining contracts are traded under collateralization, regulated by Credit Support Annex (CSA). Both cleared and CSA deals require collateral posting, as default insurance, along with its remuneration. We cannot neglect such effects.

As a consequence, once a derivative contract is closed, we must consider the cash flows associated to the collateral margining procedure. Thus, a zero-coupon contract behaves as a dividend paying asset, and its discounted price is no longer a martingale, namely
\Eq{
\frac{V^{\rm ZC}_t}{B_t} \neq \Ex{t}{\frac{V^{\rm ZC}_T}{B_T}} \,,
}%
where $V^{\rm ZC}_t$ is the zero-coupon contract's price and $B$ is the risk-free rate bank account, the numeraire of the risk neutral measure.

\subsection{A New Pricing Framework}

In order to price a financial product (for example a derivative contract), we have to discount all the cash flows occurring after the trading position is entered. We can group them as follows:
\begin{enumerate}
\item product cash flows (e.g. coupons, dividends, premiums, etc.) inclusive of hedging instruments cash flows;
\item cash flows required by the collateral margining procedure;
\item cash flows required by the funding and investing (borrowing and lending) procedures;
\item cash flows occurring upon default (close-out procedure).
\end{enumerate}%

We refer to the two names involved in the financial contract and subject to default risk as investor (also called name ``I'', usually the bank) and counterparty (also called name ``C'', for example a corporate client, but also another bank). We denote by $\tau_I$,and $\tau_C$ respectively the default times of the investor and counterparty. We fix the portfolio time horizon $T>0$, and fix the risk-neutral pricing model $(\Omega,\mathcal{G},\mathbb{Q})$, with a filtration $(\mathcal{G}_t)_{t \in [0,T]}$ such that $\tau_C$, $\tau_I$ are $(\mathcal{G}_t)_{t \in [0,T]}$-stopping times. We denote by $\Ex{t}{\cdot}$ the conditional expectation under $\mathbb{Q}$ given $\mathcal{G}_t$, and by $\Ex{\tau_i}{\cdot}$ the conditional expectation under $\mathbb{Q}$ given the stopped filtration $\mathcal{G}_{\tau_i}$. We exclude the possibility of simultaneous defaults, and define the first default event between the two parties as the stopping time
\[
\tau := \tau_C \wedge \tau_I \,.
\]
We will also consider the market sub-filtration $({\cal F}_t)_{t \ge 0}$ that one obtains implicitly by assuming a separable structure for the complete market filtration $({\cal
G}_t)_{t \ge 0}$. ${\cal G}_t$ is then generated by the pure default-free market filtration ${\cal F}_t$ and by the filtration generated by all the relevant default times monitored up to $t$ (see for example \cite{BieleckiRutkowski2002}). 

The price ${\bar V}_t$ of a derivative, inclusive of collateralized credit and debit risk (CVA and DVA), margining costs, and funding and investing costs can be derived by following \cite{Perini2011,Perini2012}, and is given by the following master equation:
\begin{eqnarray}
\label{eq:fundingpreview}
{\bar V}_t(C,F)
& = & \Ex{t}{\Pi(t,T\wedge\tau) + \gamma(t,T\wedge\tau;C) + \varphi(t,T\wedge\tau;F) } \\\nonumber
& + & \Ex{t}{\ind{t<\tau<T} D(t,\tau) \theta_\tau(C,\varepsilon) } \,,
\end{eqnarray}%
where
\begin{itemize}
\item $\Pi(t,T)$ is the sum of all discounted payoff terms in the interval $(t,T]$, without credit or debit risk, without funding costs and without collateral cash flows. In other terms, these are the financial instrument cash flows without additional risks.  
\item $\gamma(t,T;C)$ are the collateral margining costs cash flows within the interval $(t,T]$, $C$ being the collateral account.
\item $\varphi(t,T;F)$ are the funding and investing costs cash flows within such interval, $F$ being the cash account part of the hedging strategy needed for trading, and
\item $\theta_\tau(C,\varepsilon)$ is the on-default cash flow, $\varepsilon$ being the residual value of the claim being traded at default, also interpreted as the replacement cost for the deal at default (close-out amount). It is primarily this term that originates the familiar Credit and Debit Valuation Adjustments (CVA/DVA) terms.  
\end{itemize}

We also need to define $D(t,\tau)$. Notation $D(t,u)$ in general will denote the risk-neutral default-free and funding-free discount factor, given by the ratio 
\[
D(t,u) = B_t/B_u
\;,\quad
B_t = r_t B_t dt \,,
\]
where $B$ is the bank account numeraire, driven by the risk free instantaneous interest rate $r$ and associated to the risk neutral measure $\mathbb{Q}$. This rate $r$ is assumed to be $(\mathcal{F}_t)_{t \in [0,T]}$ adapted and is the key variable in all pre-crisis term structure modelling, since all other interest rates and bonds could be defined as a function of $r$, see for example \cite{BrigoMercurio2006}.

\begin{remark}{\bf (Risk-Free Rates and Risk-Neutral Measures).}
We consider $r_t$ and the related numeraire $B_t$ and measure $\mathbb{Q}$ as unobservables, and will derive the theory based on them, showing that the final equations will only depend on market observables.
\end{remark}

The margining procedure and the liquidity policy dictate respectively the dynamics of the collateral account $C_t$ and of the funding cash account $F_t$, while the close-out amount $\varepsilon_t$ is defined by the CSA that has been agreed between the deal parties. Common market procedures, as we will see later on, may link the values of such processes to the price of the product itself, transforming the previous definition of Equation \eqref{eq:fundingpreview} into a recursive equation. This feature is hidden in simplified approaches based on adding a spread to the discount curve to accommodate collateral and funding costs. A different approach is followed by \cite{Crepey2011}, who extends the usual risk-neutral evaluation framework to include many cash accounts accruing at different rates. Despite the different initial approach, a structure that is similar to our result above for the derivative price is obtained as a solution of a backward SDE.

\subsection{A Compact Formulation in Continuous Time}

The results of \cite{Perini2011,Perini2012} allow to define in an explicit way coupons and costs in terms of the collateral process $C_t$ and close-out value $\varepsilon_{\tau}$. The authors proceed by defining a sequence of margining, funding and hedging operations, and account for their costs. Such operations are taken on discrete time-grids.

Since we are going to apply the \cite{Perini2011,Perini2012} master equation to interest-rate derivatives, where collateralization usually happens on a daily basis, and where gap risk is not large, we prefer to present such results on a continuous time-grid. Furthermore, we assume that collateral re-hypothecation is allowed, as done in practice. See \cite{BrigoCapponiPallaviciniPapatheodorou} for a discussion on re-hypothecation. We write
\[
\Pi(t,u) = \int_t^u d\pi_v \, D(t,v)
\;,\quad
\gamma(t,u;C) = \int_t^u dv \, ( r_v - {\tilde c}_v ) C_v D(t,v) \,,
\]%
\begin{eqnarray*}
\varphi(t,u;F,H) &=& \int_t^u dv \, ( r_v - {\tilde f}_v ) ({\bar V}_v(C,F) - C_v ) D(t,v) \\
                &-& \int_t^u dv \, ( r_v - {\tilde h}_v ) H_v D(t,v) \,,
\end{eqnarray*}%
where $\pi_t$ is the derivative coupon process, and the collateral, funding and market rates are defined as
\[
{\tilde c}_t := c^+_t \ind{C_t>0} + c^-_t \ind{C_t<0}
\;,\quad
{\tilde f}_t := f^+_t \ind{F_t>0} + f^-_t \ind{F_t<0}
\;,\quad
{\tilde h}_t := h^+_t \ind{H_t>0} + h^-_t \ind{H_t<0} \,,
\]
with $c^\pm$ defined in the CSA contract, $f^\pm$ defined by the treasury, and $h^\pm$ equal to the growth rate of market instruments traded to implement the hedging strategy, see \cite{Perini2012} for full details. 

We can plug the above definitions into equation \eqref{eq:fundingpreview} and, by following \cite{Perini2012}, we are able to write the following continuous-time pricing equation.
\begin{eqnarray}
\label{eq:pricing}
{\bar V}_t(C;F)
& := & \int_t^T \ExT{t}{\tilde h}{ \left( \ind{u<\tau} d\pi_u + \ind{\tau\in du} \theta_u(C,\varepsilon) \right) D(t,u;{\tilde f}) } \\\nonumber
&  + & \int_t^T du \, \ExT{t}{\tilde h}{ \ind{u<\tau} ( {\tilde f}_u - {\tilde c}_u ) C_u D(t,u;{\tilde f}) }
\end{eqnarray}%
with, in case of collateral re-hypothecation,
\[
\label{eq:theta}
\theta_{\tau}(C,\varepsilon) := \varepsilon_{\tau} - \,\ind{\tau_C<\tau_I} \lgd_C (\varepsilon_{\tau} - C_{\tau^-})^+  - \,\ind{\tau_I<\tau_C} \lgd_I (\varepsilon_{\tau} - C_{\tau^-})^- \,,
\]
where we use the notation $x^- = \min(x,0)$, and where $\ind{\tau_C<\tau_I} \lgd_C (\varepsilon_{\tau} - C_{\tau^-})^+$ originates the CVA type term after collateralization, with $\ind{\tau_I<\tau_C} \lgd_I (\varepsilon_{\tau} - C_{\tau^-})^-$ originating the DVA type term instead. The above expectations are taken under a pricing measure $\mathbb{Q}^{\tilde h}$ under which the underlying risk factors grow at a rate ${\tilde h}$, and the discount factors are defined as given by
\[
D(t,T;x) := \exp\left\{-\int_t^T du\, x_u\right\} \,.
\]%

Notice that the above pricing equation \ref{eq:pricing} is not suited for explicit numerical evaluations, since the right-hand side is still depending on the derivative price via the indicators within the collateral, funding and market rates. We could resort to numerical solutions, as in \cite{Crepey2012c}, but, since our goal is pricing interest-rate derivatives, we prefer to further specialize the pricing equation for such deals.

In order to further specialize our master equation \ref{eq:pricing}, we aim to write the price as an expectation over the derivative discounted cash flows. This is achieved by adopting a few further assumptions. In particular, we assume that gap risk is not present, and we consider a particular form for collateral and close-out prices, namely
\[
C_t \doteq \alpha_t {\bar V}_t(C,F)
\;,\quad
\varepsilon_\tau \doteq {\bar V}_\tau(C,F)
\]%
with $0\le \alpha_t \le 1$. This means that
\begin{itemize}
\item[(i)] Collateral is a fraction $\alpha_t$ of the all-inclusive mark-to-market.
\item[(ii)] Close-out is the all-inclusive mark-to-market at the first default time.
\end{itemize}
This approach is considered in \cite{Biffis2011} and in the longevity-risk chapter of \cite{BrigoMoriniPallavicini2012}. An alternative approximation, not imposing a proportionality between the account pricing processes, and suited for contracts with gap risk, such as CDS, can be found in \cite{Pallavicini2013}.

We obtain, by switching to the market filtration $\cal F$,
\begin{eqnarray}
\label{eq:pricing_implicit}
{\bar V}_t(C;F)
& = & \ind{\tau>t} \int_t^T \ExT{t}{\tilde h}{ d\pi_u D(t,u;{\tilde f}+\lambda) \big| {\cal F} } \\\nonumber
& - & \ind{\tau>t} \int_t^T du\, \ExT{t}{\tilde h}{ ({\tilde\zeta}_u - \lambda_u) {\bar V}_u(C,F) D(t,u;{\tilde f}+\lambda) \big| {\cal F} } \end{eqnarray}%
with $\Ex{t}{\cdot|{\cal F}} := {\mathbb E}\!\left[\cdot|{\cal F}_t\right]$, and we define
\begin{equation}
\label{eq:zeta}
{\tilde\zeta}_t := (1 - \alpha_t) \left( \lambda^{C<I}_t \lgd_C \ind{{\bar V}_t(C;F)>0} + \lambda^{I<C}_t \lgd_I \ind{{\bar V}_t(C;F)<0} \right) - \alpha_t ( {\tilde f}_t - {\tilde c}_t ) \,,
\end{equation}%
where $\lgd_C$ and $\lgd_I$ are the loss given default respectively of the counterparty and the investor, and the default intensities are defined as
\[
\ind{\tau>t} \lambda^{C<I}_t \,dt := \Ex{t}{\ind{\tau\in dt}\ind{\tau_C<\tau_I} }
\;,\quad
\ind{\tau>t} \lambda^{I<C}_t \,dt := \Ex{t}{\ind{\tau\in dt}\ind{\tau_I<\tau_C} }
\]%
so that
\[
\ind{\tau>t} \lambda_t \,dt := \Ex{t}{\ind{\tau\in dt} } = \ind{\tau>t} (\lambda^{C<I}_t + \lambda^{I<C}_t) \,dt \,.
\]%

Hence, it is possible to apply again the Feynman-Kac theorem to obtain a differential equation, which can be solved to obtain
\begin{equation}
\label{eq:pricing_explicit}
{\bar V}_t(C;F) = \ind{\tau>t} \int_t^T \ExT{t}{\tilde h}{ d\pi_u D(t,u;{\tilde f}+{\tilde\zeta}) \big| {\cal F} } \,.
\end{equation}%
As a last step we can specialize further the above pricing equation to two extreme cases.
\begin{itemize}
\item Perfect collateralization, namely $\alpha_t=1$, and
\begin{equation}
\label{eq:collperf}
{\bar V}_t(C;F) = \ind{\tau>t} \int_t^T \ExT{t}{\tilde h}{ d\pi_u D(t,u;{\tilde c}) \big| {\cal F} } \,.
\end{equation}%
\item No collateralization, namely $\alpha_t=0$.
\begin{equation}
\label{eq:nocoll}
{\bar V}_t(C;F) = \ind{\tau>t} \int_t^T \ExT{t}{\tilde h}{ d\pi_u D(t,u;{\tilde f}+{\tilde\lambda}) \big| {\cal F} }
\end{equation}%
with
\[
{\tilde\lambda}_t = \lambda^{C<I}_t \lgd_C \ind{{\bar V}_t(C;F)>0} + \lambda^{I<C}_t \lgd_I \ind{{\bar V}_t(C;F)<0} \,.
\]%
\end{itemize}

\section{Pricing Collateralized Interest-Rate Derivatives}
\label{sec:ir}

All liquid market quotes on the money market corresponds to instruments with daily collateralization at overnight rate ($e_t$), both for the investor and the counterparty, namely
\[
{\tilde c}_t \doteq e_t \,.
\]%
Notice that the collateral accrual rate is symmetric, so that we no longer have a dependency of the accrual rates on the collateral price.

We will describe some of these instruments, such as OIS and IRS, along with their underling market rates, in the following sections. At the moment of writing this paper the market is moving from OTC contracts regulated by a (standardized) bilateral CSA to a market cleared by CCPs. We will see in Section~\ref{sec:ccp} how this change will affect the pricing of money-market instruments, while in this section we assume the bilateral view.

We use such instruments as building blocks to implement the hedging strategies for exotic interest-rate derivatives. Thus, the borrowing and lending rates ${\tilde h}_t$ will be determined according to the quotes of OIS and IRS products, which in turn depend on the overnight rate $e_t$.

In the following sections, we adopt the perfect collateralization approximation of Equation~\eqref{eq:collperf} to derive the pricing equations for OIS and IRS products, then we show how to calculate the value of the market rates ${\tilde h}_t$.

Furthermore, we assume that daily collateralization can be considered as a perfect continuous collateralization, and, in particular, we disregard gap risk. See \cite{BrigoCapponiPallaviciniPapatheodorou} for a discussion on the impact of partial collateralization on interest-rate derivatives.

\subsection{Overnight Rates and OIS}
\label{sec:ois}

The money market usually quotes the prices of collateralized instruments. In particular, a daily collateralization procedure is assumed, so that CSA contracts require that the collateral account be remunerated at the overnight rate ($e_t$). In particular, the money market usually quotes the price of Overnight Indexed Swaps (OIS). Such contracts exchange a fix-payment leg with a floating leg paying the {\emph{same}} overnight rate used for their collateralization, compounded daily. Since we are going to price OIS under the assumption of perfect collateralization, namely we are assuming that daily collateralization may be viewed as done on a continuous basis, we approximate also daily compounding in OIS floating leg with continuous compounding, which is reasonable when there is no gap risk, and we can write the one-period OIS payoff with maturity $T$ and tenor $x$ as given by
\[
1 + x K - \exp\left\{\int_{T-x}^T du\, e_u \right\}
\]%
where $K$ is the fixed rate payed by the OIS. Furthermore, we can introduce the (par) fix rates $E_t(T,x;e)$ that make the one-period OIS contract fair, namely priced at $0$. They are implicitly defined as
\[
{\bar V}^{\rm OIS}_t(K) := \ExT{t}{\tilde h}{\left( 1 + x K - \exp\left\{\int_{T-x}^T du\, e_u \right\} \right) D(t,T;e) \big| {\cal F} }
\]%
with
\[
{\bar V}^{\rm OIS}_t( E_t(T,x;e) ) = 0
\]
leading to
\[
E_t(T,x;e) := \frac{1}{x} \left( \frac{P_t(T-x;e)}{P_t(T;e)} - 1 \right)
\]%
where we define collateralized zero-coupon bonds\footnote{Notice that we are only defining a price process for hypothetical collateralized zero-coupon bond. We are not assuming that collateralized bonds are assets traded on the market.} as
\begin{equation}
\label{eq:zcbond}
P_t(T;e) := \ExT{t}{\tilde h}{D(t,T;e)\big| {\cal F} } \,.
\end{equation}%

One-period OIS rates $E_t(T,x;e)$, along with multi-period ones, are actively traded on the market. Notice that we can bootstrap collateralized zero-coupon bond prices from OIS quotes.

\begin{remark} {\bf (Hedging by means of OIS contracts).}
Exotic interest-rate products may be hedged with a strategy containing (one-period) OIS contracts. In particular, the corresponding borrowing/lending rate ${\tilde h}_t$ will be equal to the growth rate of the (one-period) OIS contract
\[
{\tilde h}_t^{\rm OIS} {\bar V}^{\rm OIS}_t  \,dt \doteq \ExT{t}{\tilde h}{d{\bar V}^{\rm OIS}_t \big| {\cal F} } = e_t {\bar V}^{\rm OIS}_t \,dt
\]
where we notice that we obtain a symmetric rate, namely the same for going long or short in OIS contracts when hedging. The extension to multi-period contracts is straightforward.
\end{remark}

\subsection{LIBOR Rates}

LIBOR rates ($L_t(T)$) are the indices used as reference rate for many collateralized interest-rate derivatives (IRS, basis swaps, \ldots). In particular we consider Interest Rate Swaps (IRS). IRS contracts swap a fix-payment leg with a floating leg paying simply compounded LIBOR rates. IRS contracts are collateralized at overnight rate $e_t$. Thus, a one-period IRS payoff with maturity $T$ and tenor $x$ is given by
\[
x K - x L_{T-x}(T)
\]%
where $K$ is the fix rate payed by the IRS. Furthermore, we can introduce the (par) fix rates $F_t(T,x;e)$ which render the one-period IRS contract fair, i.e. priced at zero. They are implicitly defined as
\[
{\bar V}^{\rm IRS}_t(K) := \ExT{t}{\tilde h}{\left( x K - x L_{T-x}(T) \right) D(t,T;e)\big| {\cal F} }
\]%
with
\[
{\bar V}^{\rm IRS}_t(F_t(T,x;e)) = 0
\]%
leading to
\[
F_t(T,x;e) := \frac{\ExT{t}{\tilde h}{L_{T-x}(T) D(t,T;e)\big| {\cal F} }}{\ExT{t}{\tilde h}{D(t,T;e)\big| {\cal F} }} = \frac{\ExT{t}{\tilde h}{L_{T-x}(T) D(t,T;e)\big| {\cal F} }}{ P_t(T;e) }
\]%

The above definition may be simplified by a suitable choice of the measure under which we take the expectation. In particular, we can consider the following Radon-Nikodym derivative
\[
\left.\frac{d \mathbb{Q}^{{\tilde h},T;e}}{d \mathbb{Q}^{\tilde h}}\right|_t := \ExT{t}{\tilde h}{D(0,T;e)\big| {\cal F} } = D(0,t;e) P_t(T;e)
\]
which is a positive $\mathbb{Q}^{\tilde h}$-martingale. We name the corresponding equivalent measure the collateralized $T$-forward measure $\ExT{t}{\tilde h,T;e}{\cdot}$.

Thus, for any payoff $\phi_T$, perfectly collateralized at overnight rate $e_t$, we can express prices as expectations under the collateralized $T$-forward measure, and we get
\[
\ExT{t}{\tilde h}{\phi_T D(t,T;e)\big| {\cal F} } = P_t(T;e) \,\ExT{t}{\tilde h,T;e}{\phi_T\big| {\cal F} }
\]%
In particular, we can write LIBOR forward rates as
\begin{equation}
\label{eq:forward}
F_t(T,x;e) := \frac{\ExT{t}{\tilde h}{L_{T-x}(T) D(t,T;e)\big| {\cal F}}}{\ExT{t}{\tilde h}{D(t,T;e)\big| {\cal F}}} = \ExT{t}{\tilde h,T;e}{L_{T-x}(T)\big| {\cal F} }
\end{equation}%

One-period forward rates $F_t(T,x;e)$, along with multi-period ones (swap rates), are actively traded on the market. Once collateralized zero-coupon bonds are derived, we can bootstrap forward rate curves from such quotes. See, for instance, \cite{Bianchetti2009} or \cite{PallaviciniTarenghi} for a discussion on bootstrapping algorithms.

\begin{remark} {\bf (Hedging by means of IRS contracts).}
Exotic interest-rate products may be hedged with a strategy containing (one-period) IRS contracts. In particular, the corresponding borrowing/lending rate ${\tilde h}_t$ will be equal to the growth rate of the IRS contract
\[
{\tilde h}_t^{\rm IRS} {\bar V}^{\rm IRS}_t \,dt \doteq \ExT{t}{\tilde h}{d{\bar V}^{\rm IRS}_t \big| {\cal F} } = e_t {\bar V}^{\rm IRS}_t \,dt
\]
where we notice that we obtain a symmetric rate, namely the same for going long or short in (one-period) IRS contracts when hedging. The extension to multi-period contracts is straightforward.
\end{remark}

\subsection{Modelling Constraints}

Our aim is to setup a multiple-curve dynamical model starting from collateralized zero-coupon bonds $P_t(T;e)$, and LIBOR forward rates $F_t(T,x;e)$. As we have seen we can bootstrap the initial curves for such quantities from the market. Now, we wish to propose a dynamics that preserves the martingale properties satisfied by such quantities.

Thus, without loss of generality, we can define collateralized zero-coupon bonds under the $\mathbb{Q}^{\tilde h}$ measure as
\[
\frac{dP_t(T;e)}{P_t(T;e)} = e_t \,dt - \alpha_t(T;e) \cdot \,dW_t^{\tilde h}
\]%
and LIBOR forward rates under $\mathbb{Q}^{{\tilde h},T;e}$ measure as
\[
dF_t(T,x;e) = \beta_t(T,x;e) \cdot \,dZ_t^{\tilde h,T;e}
\]%
where $W$s and $Z$s are correlated standard Brownian motions with correlation matrix $\rho$, and the volatility processes $\alpha$ and $\beta$ may depend on bonds and LIBOR forward rates themselves.

The following definition of $f_t(T,e)$ is not strictly necessary, and we could keep working with bonds $P_t(T;e)$, using their dynamics. However, as it is usual in interest rate theory to model rates rather than bonds, we may try to formulate quantities that are closer to the standard HJM framework. In this sense we can define instantaneous forward rates $f_t(T;e)$, by starting from (collateralized) zero-coupon bonds, as given by
\[
f_t(T;e) := -\frac{\partial}{\partial T} \log P_t(T;e)
\]%
We can derive instantaneous forward-rate dynamics by It\^o lemma, and we get under $\mathbb{Q}^{{\tilde h},T;e}$ measure
\[
df_t(T;e) = \sigma_t(T;e) \cdot \,dW_t^{\tilde h,T;e}
\;,\quad
\sigma_t(T;e) := \frac{\partial}{\partial T} \,\alpha_t(T;e)
\]%

Hence, we can summarize our modelling assumptions in the following way.
\begin{enumerate}
\item We choose to hedge exotic options in term of linear collateralized products (OIS, IRS, \ldots) which we assume to be perfectly collateralized;
\item Since such linear products can be expressed in term of simpler quantities, namely collateralized zero-coupon bonds $P_t(T;e)$ and LIBOR forward rates $F_t(T,x;e)$, we focus on their modelling. 
\item Initial term structures for collateralized products may be bootstrapped from market data. 
\item We can write rates dynamics by enforcing suitable martingale properties.
\begin{equation}
\label{eq:mm}
df_t(T;e) = \sigma_t(T;e) \cdot dW_t^{\tilde h,T;e}
\;,\quad
dF_t(T,x;e) = \beta_t(T,x;e) \cdot dZ_t^{\tilde h,T;e}
\end{equation}%
\end{enumerate}

As we explained in the introduction, this is where the multiple curve picture finally shows up: we have a curve with LIBOR based forward rates $F_t(T,x;e)$, that are collateral adjusted expectation of LIBOR market rates $L_{T-x}(T)$ we take as primitive rates from the market, and we have instantaneous forward rates $f_t(T;e)$ that are OIS based rates. OIS rates $f_t(T;e)$ are driven by collateral fees, whereas LIBOR forward rates $F_t(T,x;e)$ are driven both by collateral rates and by the primitive LIBOR market rates. 

Then, we wish to stress an important property of the above dynamics, namely that {\emph{it does not depend on unobservable rates such as the risk-free rate $r_t$}}.

Once this is done, the framework for multiple curves is ready and we may start populating the specific dynamics with modelling choices. We propose a possible choice in the following section.

\begin{remark}{\bf (Introducing a Collateral Measure).}
When we price exotic interest-rate derivatives and we hedge by means of OIS or IRS products (we can generalize the argument to other basic money-market products such as FRA or others), we obtain that the borrowing/lending rate ${\tilde h}_t$ is equal to the overnight rate $e_t$, so that we gets
\[
{\tilde h}_t \doteq {\tilde c}_t \doteq e_t \,.
\]
Hence, under the previous assumptions we can write the pricing equation~\eqref{eq:collperf} as
\[
{\bar V}_t(C;F) = \ind{\tau>t} \int_t^T \ExT{t}{e}{ d\pi_u D(t,u;e) \big| {\cal F} }
\]%
and we can speak of pricing under a ``collateralized measure''. The above equation is formally equal to the pricing equation of the Black and Scholes theory, see \cite{Piterbarg2012}, but here (i) the collateral rate is specified by the CSA contract and it can be risky; (ii) the pricing measure is introduced by means of the Feynman-Kac theorem starting from a formulation of the theory under the risk-neutral measure; (iii) the risk-free rate is not a rate which can be observed on the market. To avoid confusion we prefer not to explicitly introduce such collateralized measure.
\end{remark}

\subsection{Multiple-Curve Collateralized HJM Framework}

We can now specialize our modelling assumptions to define a model for interest-rate derivatives which is flexible enough to calibrate the quotes of the money market, but yet robust. Our aim is to to define a HJM framework to describe with a single family of Markov processes all the yield curves we are interested in. We follow \cite{Moreni2010,Moreni2012} by reformulating their theory under the $\mathbb{Q}^{\tilde h}$ measure.

In the literature many authors proposed generalizations of the HJM framework starting, see for instance \cite{Cheyette2001}, \cite{Andersen2002}, \cite{Carmona2004}, \cite{Andreasen2006}, or \cite{Chiarella2010}. In particular, in recent papers \cite{Martinez2009}, \cite{Fujii2010}, \cite{Crepey2012d} extended the HJM framework to deal with multiple-yield curves. See also \cite{Mercurio2010,Mercurio2012} for references on LIBOR Market Model.

Let us summarize the basic requirements our model must fulfill (in this section we omit the dependence on $e$ or $\tilde h$ to lighten the notation):
\begin{itemize}
\item[i)] existence of OIS rates, we can describe in term of instantaneous forward rates $f_t(T)$;
\item[ii)] existence of LIBOR rates, typical underlying of traded derivatives, with associated forwards $F_t(T,x)$;
\item[iii)] no arbitrage dynamics of the $f_t(T)$ and the $F_t(T,x)$ (both being $T$-forward measure martingales); 
\item[iv)] possibility of writing both the $f_t(T)$ and the $F_t(T,x)$ as function of a common family of Markov processes, so that we are able to build parsimonious yet flexible models.
\end{itemize}

We stress that our approach models only quantities which can bootstrapped in a model independent way from market quotes, and it includes natively the margining procedure within pricing equations. We choose under $\mathbb{Q}^{\tilde h,T;e}$ measure, the following dynamics.
\begin{eqnarray}
\label{eq:mainSDE}
df_t(T) &=& \sigma_t(T) \cdot dW_t^T \\\nonumber
\frac{dF_t(T,x)}{k(T,x)+F_t(T,x)} &=& \Sigma_t(T,x) \cdot dW_t^T
\end{eqnarray}%
where we introduce the families of (stochastic) volatility processes $\sigma_t(T)$ and $\Sigma_t(T,x),$ the vector of independent $\mathbb{Q}^{\tilde h,T;e}$-Brownian motions $W_t^T,$ and the set of deterministic shifts $k(T,x),$ such that $k(T,x)\approx 1/x$ if $x\approx 0$. We bootstrap $f_0(T)$ and $F_0(T,x)$ from market quotes.

In order to satisfy requirement iv), getting a model with a reduced number of common driving factors in the spirit of HJM approaches, it is sufficient to conveniently tie together the volatility processes $\sigma_t(T)$ and $\Sigma_t(T,x)$ as in \cite{Moreni2010}. In doing so, we extend the single-curve HJM approach of \cite{Cheyette1992}, \cite{Babbs1993}, \cite{Carverhill1994}, and \cite{Ritchken1995}. First, we consider a common parametrization for risk-free and LIBOR forward rate volatilities by introducing the family of stochastic processes $\sigma_t(u;T,x)$ such that 
\begin{equation}
\label{eq:commonVol}
\sigma_t(T) := \sigma_t(T;T,0)
\;,\quad
\Sigma_t(T,x) := \int_{T-x}^T\!\!\!\!du\, \sigma_t(u;T,x)\,.
\end{equation}
Second, we impose the separability constraint
\begin{equation}
\label{eq:separableVol}
\sigma_t(u;T,x) := v_t \cdot (q(u;T,x) g(t,u))
\;,\quad
g(t,u) := \exp\left\{-\int_t^u \!\!ds\,a(s)\right\}
\;,\quad
q(u;u,0) := 1 \,,
\end{equation}
where $v_t$ is a matrix adapted process, $q(u;T,x)$ and $a(t)$ are deterministic array functions. The condition on $q(u;T,x)$ when $T=u$ ensures that the standard HJM fulfills the usual Ritchen-Sankarasubramanian's separability condition. By plugging the expression for the volatility into Equation~\eqref{eq:mainSDE}, it is possible to work out the expression ending up with the representation
\begin{multline}\label{eq:separableLnF}
\ln\left(\frac{k(T,x)+F_t(T,x)}{k(T,x)+F_0(T,x)}\right) = \\ G^*(t,T-x,T;T,x)\cdot \left(X_t+Y_t\cdot \left(G_0(t,t,T)-\frac{1}{2}G(t,T-x,T;T,x)\right)\right) \,,
\end{multline}%
where we have defined the stochastic vector process $X_t$ and the auxiliary matrix process $Y_t$ under $\mathbb{Q}^{\tilde h}$ as given by
\begin{eqnarray*}
X_t & := & \sum_{k=1}^N \int_0^t \!\! (h_sg(s,t)) \cdot \left(dW_s + h_s\int_s^t\!\!\!dv\,g(s,v)\,ds\right) \\\nonumber
Y_t & := & \int_0^t\!\!\!ds \, (h_sg(s,t)) \cdot (h_sg(s,t))
\end{eqnarray*}%
with $X_0=0$ and $Y_0=0$, as well as the vectorial deterministic functions
\[
G_0(t,T_0,T_1):=\int_{T_0}^{T_1}\!\!\!dv\,g(t,v)
\;,\quad
G(t,T_0,T_1;T,x):= \int_{T_0}^{T_1}\!\!\!dv\,q(v;T,x)g(t,v)
\]%

It is worth noting that the integral representation of forward LIBOR volatilities given by Equation~\eqref{eq:commonVol}, together with the common separability constraint given in Equation~\eqref{eq:separableVol} are sufficient conditions to ensure the existence of a reconstruction formula for all OIS and LIBOR forward rates based on the very same family of Markov processes.

In order to model implied volatility smiles, we can add a stochastic volatility process to our model. A tractable choice is to model the matrix process $h_t$ by means of a square-root process as in \cite{Moreni2012}. In practice we set $h_t := \sqrt{v_t} R$ where $R$ is a upper triangular matrix, and the variance $v_t$ is a vector process whose dynamics under $\mathbb{Q}^{\tilde h}$ measure is given by
\[
dv_t = \kappa \left( \theta - v_t \right)\,dt + \nu\sqrt{v_t} \,dZ_t
\;,\quad
v_0 = {\bar v}
\]%
where $\kappa$,$\theta$,$\nu$,$\bar v$ are constant deterministic vectors, and $Z_t$ is a vector of independent Brownian motions correlated to the $W_t$ processes as
\[
\rho_{ij} \,dt := d\langle Z_i,W_j \rangle_t
\]%
where $\rho$ is a deterministic correlation matrix.

We address the reader for calibration and pricing examples to \cite{Moreni2010} for the cap/floor market, and to \cite{Moreni2012} for the swaption market.

\section{Collateralization Policies: from partial collateralization to CCPs}
\label{sec:funding}

In the previous section we have seen that liquid instruments quoted on the money market can be considered as fully collateralized. The HJM framework we have developed so far is based on such assumption, and it allows us to calculate prices of fully-collateralized derivative contracts. On the other hand, in practice we have often the need to evaluate uncollateralized or partially-collateralized contracts, since many counterparties, in particular corporates or smaller banks, may have difficulties to subscribe to standard CSA agreements, which require a reserve of eligible assets to be used in the margining procedure. We need also to evaluate over-collateralized deals when dealing with CCPs because of initial margin, as we shall see in the following sections.

\subsection{Pricing Non-Perfectly-Collateralized Deals}

When we trade an exotic interest-rate derivative contract we can hedge interest-rate risks by going long or short in money-market liquid instruments. On the money market the liquid contracts are usually collateralized on a daily basis at overnight rate as we have seen in Section~\ref{sec:ir}. Thus, whatever the collateralization procedure of the exotic contract is, we assume to implement the hedging strategy by means of overnight collateralized contracts. 

Hence, prices can be calculated by means of Equation~\eqref{eq:pricing_explicit}, namely we have
\[
{\bar V}_t(C;F) = \ind{\tau>t} \int_t^T \ExT{t}{\tilde h}{ d\pi_u D(t,u;{\tilde f}+{\tilde\zeta}) \big| {\cal F} }
\]%
where the risk factors grow at rate ${\tilde h}$, which is equal to the overnight rate $e_t$ when the risk factor is one of the money-market collateralized contracts (for instance the one-period OIS or IRS). Notice that, in general, the collateral accrual rate ${\tilde c}_t$, entering the pricing equation due to the definition of ${\tilde \zeta}_t$ given by Equation~\eqref{eq:zeta}, may be different from the overnight rate $e_t$.

\subsection{Convexity Adjustments for LIBOR Rates}

We can apply the pricing formula~\eqref{eq:pricing_explicit} to evaluate the money market products we introduce in the previous section, namely interest-rate swaps. Here, for example, we consider the par rate ${\bar F}_{t}(T,x;e)$ for a (partially-collateralized) single-period IRS. Such rate implicitly defined as
\[
\ind{\tau>t} \ExT{t}{\tilde h}{\left( x {\bar F}_t(T,x;e) - x L_{T-x}(T) \right) D(t,T;{\tilde f}+{\tilde \zeta}) \big| {\cal F} } = 0
\]%
The above definition can be simplified by moving to $\mathbb{Q}^{\tilde h,T;e}$ measure applying Equation~\eqref{eq:pricing_explicit}.
\[
\ind{\tau>t} \ExT{t}{{\tilde h},T;e}{\left( x {\bar F}_t(T,x;e) - x L_{T-x}(T) \right) D(t,T;{\tilde f}+{\tilde \zeta}-e) \big| {\cal F} } = 0
\]%
leading for $\tau>t$ to
\begin{equation}
{\bar F}_t(T,x;e) := \frac{ \ExT{t}{{\tilde h},T;e}{L_{T-x}(T) D(t,T;{\tilde q}) \big| {\cal F} } }{ \ExT{t}{{\tilde h},T;e}{D(t,T;{\tilde q}) \big| {\cal F} } }
\end{equation}%
where we define the effective dividend rate
\begin{equation}
{\tilde q}_t := {\tilde f}_t + {\tilde \zeta}_t - e_t
\end{equation}%
which includes the effects of credit risk, funding, and the mismatch in collateralization between the exotic deal and the instruments used in its hedging strategy.

We can now express the par rates in term of collateralized LIBOR rates, we have for $\tau>t$
\begin{eqnarray*}
{\bar F}_t(T,x;e) 
&=& \frac{ \ExT{t}{{\tilde h},T;e}{L_{T-x}(T) D(t,T;{\tilde q}) \big| {\cal F} } }{ \ExT{t}{{\tilde h},T;e}{D(t,T;{\tilde q}) \big| {\cal F} } } \\
&=& \frac{ \ExT{t}{{\tilde h},T;e}{L_{T-x}(T) \big| {\cal F} } \ExT{t}{{\tilde h},T;e}{ D(t,T;{\tilde q}) \big| {\cal F} } + \cov{t}{{\tilde h},T;e}{L_{T-x}(T) D(t,T;{\tilde q})} }{ \ExT{t}{{\tilde h},T;e}{D(t,T;{\tilde q}) \big| {\cal F} } } \\
&=& \frac{ F_t(T,x;e) \ExT{t}{{\tilde h},T;e}{ D(t,T;{\tilde q}) \big| {\cal F} } + \cov{t}{{\tilde h},T;e}{F_{T-x}(T,x;e) D(t,T;{\tilde q})} }{ \ExT{t}{{\tilde h},T;e}{D(t,T;{\tilde q}) \big| {\cal F} } }
\end{eqnarray*}%
leading to the following expression for partially collateralized LIBOR forward rates
\begin{equation}
\label{eq:forward_adj}
{\bar F}_t(T,x;e) = F_t(T,x;e) ( 1 + \gamma_t(T,x;e) )
\end{equation}%
where $\gamma_t(T,x;e)$ is the partial collateralization convexity adjustment given by
\begin{equation}
\label{eq:convexity_adj}
\gamma_t(T,x;e) := \frac{ \cov{t}{{\tilde h},T;e}{F_{T-x}(T,x;e) , D(t,T;{\tilde q}) } }{ F_t(T,x;e) \,\ExT{t}{{\tilde h},T;e}{D(t,T;{\tilde q})\big| {\cal F} } } \,.
\end{equation}%

Hence, the price of a partially-collateralized one-period IRS contract paying a fix rate $K$ can be calculated as
\begin{eqnarray}
\label{eq:irs_funded}
{\bar V}^{\rm IRS}_t
&:=& \ind{\tau>t} \ExT{t}{\tilde h}{\left( x K - x L_{T-x}(T) \right) D(t,T;{\tilde f}+{\tilde \zeta})\big| {\cal F} } \\\nonumber
& =& \ind{\tau>t} \,x \left(K - {\bar F}_{t}(T,x;e) \right) {\bar P}_t(T;e) 
\end{eqnarray}%
where we have defined for convenience for $\tau>t$ the (partially) collateralized adjusted zero-coupon bond as
\begin{equation}
\label{eq:zcbond_adj}
{\bar P}_t(T;e) := P_t(T;e) \,\ExT{t}{\tilde h,T;e}{D(t,T;{\tilde q}) \big| {\cal F} }
\end{equation}%

These results can be straightforwardly extended to multi-period contracts.

Notice that partially collateralized zero-coupon bonds and forward rates depend on the price process of the contract paying them. Thus, they have different values for different contracts. We can interpret them respectively as a {\emph{per-contract}} $Z$-spread-adjusted bonds and convexity-adjusted forward rates. Thus, when collateralization is not perfect, we obtain that each contract has its own curve.

\subsection{Initial Margins and CCPs}
\label{sec:ccp}

Funding costs may arise also in collateralized contract, if they are closed with a Central Clearing Counterparty (CCP). Indeed, CCP requires that counterparties post an initial margin to close the deal. Along with the initial margin CCPs require also regular posting of collaterals (variation margin) to match the mark-to-market variation of the deal. The analysis of the impact of variation margin procedures can be found in \cite{Rama2011}, where convexity adjustments and NPV effects are discussed for different clearing houses. These two effects, in the case of one-period contracts, correspond respectively to the adjustments of LIBOR forward rates, given in Equation~\eqref{eq:forward_adj}, and to the adjustment in zero-coupon bonds, given in Equation~\eqref{eq:zcbond_adj}. In this section, we focus on initial margins and their contribution to funding costs.

Initial margins are collected to hedge potential future counterparty exposures resulting in replacement costs. In particular, a CCP is vulnerable to losses on defaulting counterparty exposures between the time of the last variation margin payment of the defaulting counterparty and close-out valuation. This is known as margin period of risk, see \cite{Heller2012}. Furthermore, initial margin may cover gap risks. Initial margins are calculated at contract inception, and, then, possibly updated during the deal life-time. The amount of initial margin can be estimated by a CCP according to Value-at-Risk or expected shortfall risk measure.

We can deal with initial margin by assuming that an over-collateralization is required by the contract, namely we can generalize the argument leading to Equation\eqref{eq:pricing_implicit} to allow for $\alpha_t>1$. By a direct calculation starting from Equation~\eqref{eq:pricing} we obtain
\begin{equation}
\label{eq:pricing_explicit_ext}
{\bar V}_t(C;F) = \ind{\tau>t} \int_t^T \ExT{t}{\tilde h}{ d\pi_u D(t,u;{\tilde f}+{\tilde\xi}) \big| {\cal F} }
\end{equation}
where we define
\begin{eqnarray}
\label{eq:xi}
{\tilde\xi}_t
&:=& (1 - \alpha_t)^+ \left( \lambda^{C<I}_t \lgd_C \ind{{\bar V}_t(C;F)>0} + \lambda^{I<C}_t \lgd_I \ind{{\bar V}_t(C;F)<0} \right) \\\nonumber
& +& (1 - \alpha_t)^- \left( \lambda^{I<C}_t \lgd_I \ind{{\bar V}_t(C;F)>0} + \lambda^{C<I}_t \lgd_C \ind{{\bar V}_t(C;F)<0} \right) \\\nonumber
& -& \alpha_t ( {\tilde f}_t - {\tilde c}_t )
\end{eqnarray}%
In the case $0\le\alpha_t\le1$ we get back Equation~\eqref{eq:pricing_explicit_ext} with ${\tilde\xi}_t$ replaced by ${\tilde\zeta}_t$.

In order to interpret the above result, we consider with a closer look the differences between the partial, the perfect and the CCP collateralization cases.
\begin{itemize}
\item Partial collateralization, namely $0\le\alpha_t<1$.
\[
{\tilde\xi}_t = (1 - \alpha_t) \left( \lambda^{C<I}_t \lgd_C \ind{{\bar V}_t(C;F)>0} + \lambda^{I<C}_t \lgd_I \ind{{\bar V}_t(C;F)<0} \right) - \alpha_t ( {\tilde f}_t - {\tilde c}_t )
\]%
The effective rate ${\tilde\xi}_t$ has three contributions: (i) a CVA term accounting for the non-collateralized fraction of exposure, if positive; (ii) an analogous DVA term when exposure is negative; (iii) a term accounting for collateral funding costs.
\item Perfect collateralization, namely $\alpha_t=1$.
\[
{\tilde\xi}_t = {\tilde f}_t - {\tilde c}_t
\]%
The effective rate ${\tilde\xi}_t$ has only one contribution: (i) a term accounting for collateral funding costs.
\item CCP collateralization, namely $\alpha_t>1$.
\[
{\tilde\xi}_t = (1 - \alpha_t) \left( \lambda^{I<C}_t \lgd_I \ind{{\bar V}_t(C;F)>0} + \lambda^{C<I}_t \lgd_C \ind{{\bar V}_t(C;F)<0} \right) - \alpha_t ( {\tilde f}_t - {\tilde c}_t )
\]%
The effective rate ${\tilde\xi}_t$ has three contributions: (i) a DVA term accounting for the excess of collateral to be given back to the CCP, if exposure is positive; (ii) an analogous CVA term when exposure is negative (which can be discarded if we assume CCP being default free); (iii) a term accounting for collateral funding costs. Notice than the first two term are due to re-hypothecation, namely to the fact that the surviving party must resort to an unsecured claim to get back the collateral excess. Indeed, if we repeat the derivation of Equation \eqref{eq:pricing_explicit_ext} without the re-hypothecation hypothesis we obtain only the third contribution.

\end{itemize}

\section{Modelling Credit Spreads, Funding Rates and Collateral Fractions}
\label{sec:modelling}

When pricing a partially-, non- or over-collateralized deal we have to define credit spreads and funding rates. In the following sections we suggest how to write their dynamics. Our intention is to introduce simple yet realistic solutions which can be used to implement numerical calculations.

\subsection{Credit Spreads and Gap Risk}

The pricing equation is dependent on the future prices of deal, since CVA/DVA and investing/funding costs depend on their sign, so that we must resort to numerical tools to solve the backward induction needed to calculate the price.

Yet, this is not the greatest problem with non-perfectly-collateralized deals. Indeed, {\emph{the real problem is that the market is incomplete}}. The term structure of credit spreads is usually derived from CDS quotes, but we have no, or very few, informations from option markets. Moreover, we should introduce model-dependent hypotheses as soon as we consider gap risk into CDS evaluation to describe co-dependency of the default events for the counterparties and the reference name of the CDS, as described in \cite{BrigoCapponiPallavicini}. Furthermore, gap risk prevent from considering CDS contracts as fully collateralized, so that we should introduce funding costs from the beginning of the calibration procedure.

In the practice such considerations are ignored and credit spreads are calibrated to CDS contracts by considering them as fully-collateralized contracts, so that we could define the credit-spread dynamics without introducing funding costs, along with some specification to model the co-dependency between default events. Yet, the impact of such choice is to underestimate the default probabilities coming from CDS quotes. For instance, in \cite{Fujii2011CDS} for stylized market settings the authors find a correction due to gap risk on CDS spreads up to $10\%$ at short maturities, and up to $50\%$ at long maturities. Such uncertainty should be compared to the uncertainty in recovery rates.

\subsection{Funding Rates, Liquidity Policies and Collateral Portfolio}

Even more problematic is the funding rate modelling.

Funding rates are determined by the Treasury department according to the Bank funding policies. Thus, a term structure of funding rates is known, but it is far from being unambiguously derived from market quotes. For instance, long maturities in the term structure of funding rates are calculated by rolling over short-term funding positions, and not by entering into a long-term funding position. It is very difficult to forecast the future strategies followed by the Treasury. Thus, the term structure of funding rates is model-depend. The option market (e.g. contingent funding derivatives) is missing.

A tempting possibility is using the LIBOR rates as a proxy of funding rates. This choice is widely spread, but it is very problematic, since it implies that the funding policies of the Treasury department is based on inter-bank deposits (not to speak of possible frauds in LIBOR published rates). After the crisis only a small part of funding comes from this source: the main source of funding for an investment Bank is the collateral portfolio which is mainly driven by the credit spreads of the underlying names.

Once a sensible model is defined for the funding policy we should consider that the Treasury department may implement a maturity transformation policy along with a fund transfer price (FTP) process. Such procedures may alter in a significant way the funding rate.

Here, we wish to select a sensible choice for the dynamics of funding and investing rates to perform numerical simulations. A possibility is to use for the funding rate $f^+_t$ an affine combination of the investor's default intensity $\lambda^I_t$ and the average default intensity $\lambda^P_t$ of a pool of financial names, representing the names underlying the collateral portfolio. Such an average can be obtained from CDS quotes or from index proxies, such as i-Traxx financial sub-index. While for the investing rate $f^-_t$ we can use a different affine combination which includes only $\lambda^P_t$. Thus, we can define the funding and investing rates as given by
\begin{equation}
\label{eq:investing_rate}
f^-_t := e_t + w^-(t) + w^P(t) \lambda^P_t
\end{equation}%
and
\begin{equation}
\label{eq:funding_rate}
f^+_t := e_t + w^+(t) + w^P(t) \lambda^P_t + w^I(t) \lambda^I_t
\end{equation}%
where $e_t$ is the overnight rate, and the $w$'s are deterministic functions of times, which can be calibrated to Treasury data and bond/CDS basis. Moreover, the $w$'s define in an implicit way the correlation between the funding/investing rates, the overnight rate and the default intensities.

\subsection{Collateral Fractions, Haircuts and Settlement Liquidity Risk}

The last unknown process to be modelled is the collateral fraction. We consider three realistic cases: (i) no collateralization, (ii) perfect collateralization, and (iii) CCP collateralization with initial margin calculated with VaR risk measure. The first two cases are simply obtained by setting respectively $\alpha_t \doteq 0$ and $\alpha_t \doteq 1$. The third case is more elaborated, since we should assume
\begin{eqnarray*}
C_t := \alpha_t {\bar V}_t(C;F) 
&\doteq&
   {\bar V}_t(C;F) \\
&+& \left( -Q_{-{\bar V}_t(C;F)}(q) \right)^+ \ind{{\bar V}_t(C;F)>\left( -Q_{-{\bar V}_t(C;F)}(q) \right)^+} \\
&+& \left( Q_{{\bar V}_t(C;F)}(q) \right)^- \ind{{\bar V}_t(C;F)<\left( Q_{{\bar V}_t(C;F)}(q) \right)^-}
\end{eqnarray*}
leading to
\begin{equation}
\label{eq:haircut_var}
\alpha_t \doteq 1 + \varsigma^+_t(q) + \varsigma^-_t(q) 
\end{equation}%
where we introduce the collateral haircuts $\varsigma^\pm_t(q)\in[0,1)$ as given by
\begin{equation}
\varsigma^\pm_t(q) := \mp \frac{(Q_{\mp{\bar V}_t(C;F)}(q))^-}{{\bar V}_t(C;F)} \ind{\mp{\bar V}_t(C;F)<(Q_{\mp{\bar V}_t(C;F)}(q))^-}
\end{equation}%
while the quantile function $Q$ for a random variable $X$ at a particular level $q$ is defined as 
\[
Q_{X}(q) := \inf\{x:q<\Px{X<x}\}
\]%
and $\mathbb{P}$ is the physical probability.

Yet, this approach can be difficult to follow within a risk-neutral pricing framework. Thus, we suggest an alternative approach which focuses on prices instead of considering probabilities. Such approach can be used to gauge the impact of funding the initial margin, but to exactly match the algorithm used by the CCP we should resort to Equation~\eqref{eq:haircut_var}.

The meaning of the haircuts $\varsigma^\pm_t(q)$, introduced starting from quantiles, is to cover the surviving party from settlement liquidity risk, namely from the risk that the replacement deal will be close some days after the default event, so that its mark-to-market may be different from the loss evaluated at the default event.

Yet, this protection can be easily achieved by means of (non-collateralized) call-like options. Indeed, we can write:
\begin{eqnarray*}
C_t &\doteq& \ind{{\bar V}_t(C;F)>0} \ExT{t}{\tilde h}{ \max\{ {\bar V}_t(C;F) , {\bar V}_{t+\delta}(C;F) D(t,t+\delta;\tilde f) \} } \\
    & +& \ind{{\bar V}_t(C;F)<0} \ExT{t}{\tilde h}{ \min\{ {\bar V}_t(C;F) , {\bar V}_{t+\delta}(C;F) D(t,t+\delta;\tilde f) \} } \\
    & =& {\bar V}_t(C;F) \left( 1 + \ExT{t}{\tilde h}{ \left( \frac{{\bar V}_{t+\delta}(C;F) D(t,t+\delta;\tilde f)}{{\bar V}_t(C;F)} - 1 \right)^+ } \right)
\end{eqnarray*}%
where $\delta$ is the margin period of risk, see \cite{BrigoCapponiPallaviciniPapatheodorou}, and can range from few days up to 20 days. In term of collateral fraction we obtain
\begin{equation}
\alpha_t \doteq 1 + \varsigma_t
\end{equation}%
where we introduce the collateral haircut $\varsigma_t>0$ as given by
\[
\varsigma_t := \ExT{t}{\tilde h}{ \left( \frac{{\bar V}_{t+\delta}(C;F) D(t,t+\delta;\tilde f)}{{\bar V}_t(C;F)} - 1 \right)^+ }
\]%
Then, in order to avoid non-realistic values for the collateral haircut, and to mimic the quantile-dependent version, we choose the following definition which ensures $\varsigma\in[0,1)$.
\begin{equation}
\label{eq:haircut}
\varsigma_t := 1 + \left( \ExT{t}{\tilde h}{ \left( \frac{{\bar V}_{t+\delta}(C;F) D(t,t+\delta;\tilde f)}{{\bar V}_t(C;F)} - 1 \right)^+ } - 1 \right)^-
\end{equation}

\section{Conclusions and Further Developments}
\label{sec:conclusion}

We model multiple LIBOR and OIS based interest rate curves consistently, based only on market observables and by consistently including credit, collateral and funding effects. Further work includes a more realistic model for initial margins in CCP's, extending our analysis to gap risk, collateralization in different currencies, and more realistic collateral processes inclusive of minimum transfer amounts, thresholds and haircuts. Also, cases where collateral is not cash can be considered.

\section*{Acknowledgments}

We are grateful to Marco Bianchetti, Antonio Castagna, Marc Henrard, Nicola Moreni, Daniele Perini and Giulio Sartorelli for helpful discussions.

\newpage

\bibliographystyle{plainnat}
\bibliography{hjm_collateral}

\begin{thebibliography}{48}
\providecommand{\natexlab}[1]{#1}
\providecommand{\url}[1]{\texttt{#1}}
\expandafter\ifx\csname urlstyle\endcsname\relax
  \providecommand{\doi}[1]{doi: #1}\else
  \providecommand{\doi}{doi: \begingroup \urlstyle{rm}\Url}\fi

\bibitem[Ametrano and Bianchetti(2009)]{Bianchetti2009}
F.~M. Ametrano and M.~Bianchetti.
\newblock Bootstrapping the illiquidity: Multiple yield curves construction for
  market coherent forward rates estimation.
\newblock In F.~Mercurio, editor, \emph{Modeling Interest Rates: Latest
  Advances for Derivatives Pricing}. Risk Books, 2009.

\bibitem[Andersen and Andreasen(2002)]{Andersen2002}
L.~Andersen and J.~Andreasen.
\newblock Volatile volatilities.
\newblock \emph{Risk Magazine}, 12, 2002.

\bibitem[Andreasen(2006)]{Andreasen2006}
J.~Andreasen.
\newblock Stochastic volatility for real.
\newblock \emph{ssrn.com}, 2006.

\bibitem[Arnsdorf(2011)]{Arnsdorf2011}
M.~Arnsdorf.
\newblock Central counterparty risk.
\newblock \emph{arXiv}, 2011.

\bibitem[Babbs(1993)]{Babbs1993}
S.~Babbs.
\newblock Generalised vasicek models of the term structure.
\newblock \emph{Applied Stochastic Models and Data Analysis}, 1:\penalty0
  46--92, 1993.

\bibitem[Bielecki and Rutkowski(2002)]{BieleckiRutkowski2002}
T.~Bielecki and M.~Rutkowski.
\newblock \emph{Credit Risk: Modeling, Valuation and Hedging.}
\newblock Springer Finance, Berlin, 2002.

\bibitem[Biffis et~al.(2011)Biffis, Blake, Pitotti, and Sun]{Biffis2011}
E.~Biffis, D.~P. Blake, L.~Pitotti, and A.~Sun.
\newblock The cost of counterparty risk and collateralization in longevity
  swaps.
\newblock 2011.

\bibitem[Brigo and Mercurio(2006)]{BrigoMercurio2006}
D.~Brigo and F.~Mercurio.
\newblock \emph{Interest Rate Models: Theory and Practice with Smile, Inflation
  and Credit}.
\newblock Springer Verlag, 2006.

\bibitem[Brigo and Pallavicini(2007)]{BrigoPallavicini2007}
D.~Brigo and A.~Pallavicini.
\newblock Counterparty risk under correlation between default and interest
  rates.
\newblock In J.~Miller, D.~Edelman, and J.~Appleby, editors, \emph{Numerical
  Methods for Finance}. Chapman Hall, 2007.

\bibitem[Brigo et~al.(2011{\natexlab{a}})Brigo, Capponi, and
  Pallavicini]{BrigoCapponiPallavicini}
D.~Brigo, A.~Capponi, and A.~Pallavicini.
\newblock Arbitrage-free bilateral counterparty risk valuation under
  collateralization and re-hypothecation with application to {CDS}.
\newblock \emph{Mathematical Finance}, 2011{\natexlab{a}}.
\newblock Accepted for publication.

\bibitem[Brigo et~al.(2011{\natexlab{b}})Brigo, Capponi, Pallavicini, and
  Papatheodorou]{BrigoCapponiPallaviciniPapatheodorou}
D.~Brigo, A.~Capponi, A.~Pallavicini, and V.~Papatheodorou.
\newblock Collateral margining in arbitrage-free counterparty valuation
  adjustment including re-hypotecation and netting.
\newblock \emph{Working Paper}, 2011{\natexlab{b}}.
\newblock URL \url{ssrn.com}.

\bibitem[Brigo et~al.(2012)Brigo, Morini, and
  Pallavicini]{BrigoMoriniPallavicini2012}
D.~Brigo, M.~Morini, and A.~Pallavicini.
\newblock \emph{Counterparty Credit Risk, Collateral and Funding with pricing
  cases for all asset classes}.
\newblock Wiley, Forthcoming, 2012.

\bibitem[Brunnermeier and Pedersen(2009)]{Brunnermeier2009}
M.~Brunnermeier and L.~Pedersen.
\newblock Market liquidity and funding liquidity.
\newblock \emph{The Review of Financial Studies}, 22 (6), 2009.

\bibitem[Burgard and Kjaer(2011{\natexlab{a}})]{BurgardKjaer2011a}
C.~Burgard and M.~Kjaer.
\newblock Partial differential equation representations of derivatives with
  counterparty risk and funding costs.
\newblock \emph{The Journal of Credit Risk}, 7 (3):\penalty0 1--19,
  2011{\natexlab{a}}.
\newblock URL \url{http://ssrn.com/abstract=1605307}.

\bibitem[Burgard and Kjaer(2011{\natexlab{b}})]{BurgardKjaer2011b}
C.~Burgard and M.~Kjaer.
\newblock In the balance.
\newblock \emph{Risk Magazine}, October, 2011{\natexlab{b}}.

\bibitem[Carmona(2004)]{Carmona2004}
R.~Carmona.
\newblock Hjm: a unified approach to dynamic models for fixed income, credit
  and equity markets.
\newblock In R.A. Carmona, E.~\c{C}inlar, I.~Ekeland, E.~Jouini, J.A.
  Scheinkman, and N.~Touzi, editors, \emph{Paris-Princeton Lectures on
  Mathematical Finance}. Springer Verlag, 2004.

\bibitem[Carverhill(1994)]{Carverhill1994}
A.~Carverhill.
\newblock When is the short rate markovian.
\newblock \emph{Mathematical Finance}, 2:\penalty0 135--154, 1994.

\bibitem[Cheyette(1992)]{Cheyette1992}
O.~Cheyette.
\newblock Term structure dynamics and mortgage valuation.
\newblock \emph{Journal of Fixed Income}, 1, 1992.

\bibitem[Cheyette(2001)]{Cheyette2001}
O.~Cheyette.
\newblock Markov representation of the heath-jarrow-morton model.
\newblock \emph{ssrn.com}, 2001.

\bibitem[Chiarella et~al.(2010)Chiarella, Maina, and
  Nikitipoulos~Sklibosios]{Chiarella2010}
C.~Chiarella, S.C. Maina, and C.~Nikitipoulos~Sklibosios.
\newblock Markovian defaultable hjm term structure models with unspanned
  stochastic volatility.
\newblock \emph{ssrn.com}, 2010.

\bibitem[Cont et~al.(2011)Cont, Mondescu, and Yu]{Rama2011}
R.~Cont, R.P. Mondescu, and Y.~Yu.
\newblock Central clearing of interest rate swaps: A comparison of offerings.
\newblock \emph{ssrn.com}, 2011.

\bibitem[Cr\'epey(2011)]{Crepey2011}
S.~Cr\'epey.
\newblock A {BSDE} approach to counterparty risk under funding constraints.
\newblock \emph{Working Paper}, 2011.
\newblock URL \url{grozny.maths.univ-evry.fr/pages\_perso/crepey}.

\bibitem[Cr\'epey(2012{\natexlab{a}})]{Crepey2012a}
S.~Cr\'epey.
\newblock Bilateral counterparty risk under funding constraints – {P}art {I}:
  {P}ricing.
\newblock \emph{Forthcoming in Mathematical Finance}, 2012{\natexlab{a}}.

\bibitem[Cr\'epey(2012{\natexlab{b}})]{Crepey2012b}
S.~Cr\'epey.
\newblock Bilateral counterparty risk under funding constraints – {P}art
  {II}: {CVA}.
\newblock \emph{Forthcoming in Mathematical Finance}, 2012{\natexlab{b}}.

\bibitem[Cr\'epey et~al.(2012{\natexlab{a}})Cr\'epey, Gerboud, Grbac, and
  Ngor]{Crepey2012c}
S.~Cr\'epey, R.~Gerboud, Z.~Grbac, and N.~Ngor.
\newblock Counterparty risk and funding: The four wings of the {TVA}.
\newblock \emph{Working Paper}, 2012{\natexlab{a}}.
\newblock URL \url{grozny.maths.univ-evry.fr/pages\_perso/crepey}.

\bibitem[Cr\'epey et~al.(2012{\natexlab{b}})Cr\'epey, Grbac, and
  Nguyen]{Crepey2012d}
S.~Cr\'epey, Z.~Grbac, and H.N. Nguyen.
\newblock A multiple-curve hjm model of interbank risk.
\newblock \emph{Mathematics and Financial Economics}, 6:\penalty0 155--190,
  2012{\natexlab{b}}.

\bibitem[Eisenschmidt and Tapking(2009)]{Tapking2009}
J.~Eisenschmidt and J.~Tapking.
\newblock Liquidity risk premia in unsecured interbank money markets.
\newblock \emph{Working Paper Series European Central Bank}, 2009.

\bibitem[Filipovic and Trolle(2012)]{Filipovic2012}
D.~Filipovic and A.B. Trolle.
\newblock The term structure of interbank risk.
\newblock \emph{ssrn.com}, 2012.

\bibitem[Fujii and Takahashi(2011{\natexlab{a}})]{Fujii2011}
M.~Fujii and A.~Takahashi.
\newblock Clean valuation framework for the usd silo.
\newblock \emph{Working Paper}, 2011{\natexlab{a}}.
\newblock URL \url{ssrn.com}.

\bibitem[Fujii and Takahashi(2011{\natexlab{b}})]{Fujii2011CDS}
M.~Fujii and A.~Takahashi.
\newblock Collateralized cds and default dependence.
\newblock \emph{Working Paper}, 2011{\natexlab{b}}.
\newblock URL \url{ssrn.com}.

\bibitem[Fujii et~al.(2010)Fujii, Shimada, and Takahashi]{Fujii2010}
M.~Fujii, Y.~Shimada, and A.~Takahashi.
\newblock Collateral posting and choice of collateral currency.
\newblock \emph{Working Paper}, 2010.
\newblock URL \url{ssrn.com}.

\bibitem[Heller and Vause(2012)]{Heller2012}
D.~Heller and N.~Vause.
\newblock From turmoil to crisis: Dislocations in the fx swap market.
\newblock \emph{BIS Working Paper}, 373, 2012.

\bibitem[Henrard(2009)]{Henrard2009}
M.~Henrard.
\newblock The irony in the derivatives discounting part ii: The crisis.
\newblock \emph{ssrn}, 2009.

\bibitem[Kijima et~al.(2009)Kijima, Tanaka, and Wong]{Kijima2009}
M.~Kijima, K.~Tanaka, and T.~Wong.
\newblock A multi-quality model of interest rates.
\newblock \emph{Quantitative Finance}, 9:\penalty0 133--145, 2009.

\bibitem[Mart{\`\i}nez(2009)]{Martinez2009}
T.~Mart{\`\i}nez.
\newblock Drift conditions on a hjm model with stochastic basis spreads.
\newblock 2009.
\newblock URL \url{www.risklab.es/es/jornadas/2009/index.html}.

\bibitem[Mercurio(2009)]{Mercurio2009}
F.~Mercurio.
\newblock Interest rates and the credit crunch: New formulas and market models.
\newblock \emph{Bloomberg Portfolio Research Paper}, 2009.

\bibitem[Mercurio(2010)]{Mercurio2010}
F.~Mercurio.
\newblock Libor market models with stochastic basis.
\newblock \emph{Risk Magazine}, 12, 2010.

\bibitem[Mercurio and Xie(2012)]{Mercurio2012}
F.~Mercurio and Z.~Xie.
\newblock The basis goes stochastic.
\newblock \emph{Risk Magazine}, 12, 2012.

\bibitem[Moreni and Pallavicini(2010)]{Moreni2010}
N.~Moreni and A.~Pallavicini.
\newblock Parsimonious hjm modelling for multiple yield-curve dynamics.
\newblock \emph{Submitted to Quantitative Finance}, 2010.

\bibitem[Moreni and Pallavicini(2012)]{Moreni2012}
N.~Moreni and A.~Pallavicini.
\newblock Parsimonious multi-curve hjm modelling with stochastic volatility.
\newblock In M~Bianchetti and M.~Morini, editors, \emph{Interest Rate Modelling
  After The Financial Crisis}. Risk Books, 2012.

\bibitem[Pallavicini(2013)]{Pallavicini2013}
A.~Pallavicini.
\newblock A comprehensive framework for bilateral collateralized cva and
  funding costs.
\newblock In \emph{Finance and Stochastics Seminar}. Imperial College, 2013.
\newblock Presented at the conference in London, January 16.

\bibitem[Pallavicini and Tarenghi(2010)]{PallaviciniTarenghi}
A.~Pallavicini and M.~Tarenghi.
\newblock Interest-rate modelling with multiple yield curves.
\newblock 2010.

\bibitem[Pallavicini et~al.(2011)Pallavicini, Perini, and Brigo]{Perini2011}
A.~Pallavicini, D.~Perini, and D.~Brigo.
\newblock Funding {V}aluation {A}djustment: {FVA} consistent with {CVA}, {DVA},
  {WWR}, {C}ollateral, {N}etting and re-hyphotecation.
\newblock \emph{arxiv.org, ssrn.com}, 2011.

\bibitem[Pallavicini et~al.(2012)Pallavicini, Perini, and Brigo]{Perini2012}
A.~Pallavicini, D.~Perini, and D.~Brigo.
\newblock Funding, {C}ollateral and {H}edging: uncovering the mechanics and the
  subtleties of funding valuation adjustments.
\newblock \emph{arxiv.org, ssrn.com}, 2012.

\bibitem[Pirrong(2011)]{Pirrong2011}
C.~Pirrong.
\newblock The economics of central clearing: Theory and practice.
\newblock \emph{ISDA Discussion Papers Series}, 2011.

\bibitem[Piterbarg(2010)]{Piterbarg2010}
V.~Piterbarg.
\newblock Funding beyond discounting: collateral agreements and derivatives
  pricing.
\newblock \emph{Risk Magazine}, 2:\penalty0 97--102, 2010.

\bibitem[Piterbarg(2012)]{Piterbarg2012}
V.~Piterbarg.
\newblock Cooking with collateral.
\newblock \emph{Risk Magazine}, 8, 2012.

\bibitem[Ritchken and Sankarasubramanian(1995)]{Ritchken1995}
P.~Ritchken and L.~Sankarasubramanian.
\newblock Volatility structures of forward rates and the dynamics of the term
  structure.
\newblock \emph{Mathematical Finance}, 7:\penalty0 157--176, 1995.

\end{thebibliography}

\end{document}